\RequirePackage{lineno}
\documentclass[aps,prc,twocolumn,superscriptaddress,showpacs]{revtex4}
\usepackage{color}
\usepackage{subfigure}
\usepackage{amsmath,epsfig}
\usepackage{graphicx}
\usepackage{dcolumn}
\usepackage{bm}
\usepackage{multirow}
\usepackage{hhline}
\begin{document}


\title{System size dependence of transverse momentum correlations at
  $\sqrt{s_{NN}} =$62.4 and 200~GeV at the BNL Relativistic Heavy Ion Collider}

\affiliation{AGH University of Science and Technology, Cracow, Poland}
\affiliation{Argonne National Laboratory, Argonne, Illinois 60439, USA}
\affiliation{University of Birmingham, Birmingham, United Kingdom}
\affiliation{Brookhaven National Laboratory, Upton, New York 11973, USA}
\affiliation{University of California, Berkeley, California 94720, USA}
\affiliation{University of California, Davis, California 95616, USA}
\affiliation{University of California, Los Angeles, California 90095, USA}
\affiliation{Universidade Estadual de Campinas, Sao Paulo, Brazil}
\affiliation{Central China Normal University (HZNU), Wuhan 430079, China}
\affiliation{University of Illinois at Chicago, Chicago, Illinois 60607, USA}
\affiliation{Cracow University of Technology, Cracow, Poland}
\affiliation{Creighton University, Omaha, Nebraska 68178, USA}
\affiliation{Czech Technical University in Prague, FNSPE, Prague, 115 19, Czech Republic}
\affiliation{Nuclear Physics Institute AS CR, 250 68 \v{R}e\v{z}/Prague, Czech Republic}
\affiliation{University of Frankfurt, Frankfurt, Germany}
\affiliation{Institute of Physics, Bhubaneswar 751005, India}
\affiliation{Indian Institute of Technology, Mumbai, India}
\affiliation{Indiana University, Bloomington, Indiana 47408, USA}
\affiliation{Alikhanov Institute for Theoretical and Experimental Physics, Moscow, Russia}
\affiliation{University of Jammu, Jammu 180001, India}
\affiliation{Joint Institute for Nuclear Research, Dubna, 141 980, Russia}
\affiliation{Kent State University, Kent, Ohio 44242, USA}
\affiliation{University of Kentucky, Lexington, Kentucky, 40506-0055, USA}
\affiliation{Institute of Modern Physics, Lanzhou, China}
\affiliation{Lawrence Berkeley National Laboratory, Berkeley, California 94720, USA}
\affiliation{Massachusetts Institute of Technology, Cambridge, MA 02139-4307, USA}
\affiliation{Max-Planck-Institut f\"ur Physik, Munich, Germany}
\affiliation{Michigan State University, East Lansing, Michigan 48824, USA}
\affiliation{Moscow Engineering Physics Institute, Moscow Russia}
\affiliation{National Institute of Science Education and Research, Bhubaneswar 751005, India}
\affiliation{Ohio State University, Columbus, Ohio 43210, USA}
\affiliation{Old Dominion University, Norfolk, VA, 23529, USA}
\affiliation{Institute of Nuclear Physics PAN, Cracow, Poland}
\affiliation{Panjab University, Chandigarh 160014, India}
\affiliation{Pennsylvania State University, University Park, Pennsylvania 16802, USA}
\affiliation{Institute of High Energy Physics, Protvino, Russia}
\affiliation{Purdue University, West Lafayette, Indiana 47907, USA}
\affiliation{Pusan National University, Pusan, Republic of Korea}
\affiliation{University of Rajasthan, Jaipur 302004, India}
\affiliation{Rice University, Houston, Texas 77251, USA}
\affiliation{Universidade de Sao Paulo, Sao Paulo, Brazil}
\affiliation{University of Science \& Technology of China, Hefei 230026, China}
\affiliation{Shandong University, Jinan, Shandong 250100, China}
\affiliation{Shanghai Institute of Applied Physics, Shanghai 201800, China}
\affiliation{SUBATECH, Nantes, France}
\affiliation{Temple University, Philadelphia, Pennsylvania, 19122}
\affiliation{Texas A\&M University, College Station, Texas 77843, USA}
\affiliation{University of Texas, Austin, Texas 78712, USA}
\affiliation{University of Houston, Houston, TX, 77204, USA}
\affiliation{Tsinghua University, Beijing 100084, China}
\affiliation{United States Naval Academy, Annapolis, MD 21402, USA}
\affiliation{Valparaiso University, Valparaiso, Indiana 46383, USA}
\affiliation{Variable Energy Cyclotron Centre, Kolkata 700064, India}
\affiliation{Warsaw University of Technology, Warsaw, Poland}
\affiliation{University of Washington, Seattle, Washington 98195, USA}
\affiliation{Wayne State University, Detroit, Michigan 48201, USA}
\affiliation{Yale University, New Haven, Connecticut 06520, USA}
\affiliation{University of Zagreb, Zagreb, HR-10002, Croatia}

\author{L.~Adamczyk}\affiliation{AGH University of Science and Technology, Cracow, Poland}
\author{J.~K.~Adkins}\affiliation{University of Kentucky, Lexington, Kentucky, 40506-0055, USA}
\author{G.~Agakishiev}\affiliation{Joint Institute for Nuclear Research, Dubna, 141 980, Russia}
\author{M.~M.~Aggarwal}\affiliation{Panjab University, Chandigarh 160014, India}
\author{Z.~Ahammed}\affiliation{Variable Energy Cyclotron Centre, Kolkata 700064, India}
\author{I.~Alekseev}\affiliation{Alikhanov Institute for Theoretical and Experimental Physics, Moscow, Russia}
\author{J.~Alford}\affiliation{Kent State University, Kent, Ohio 44242, USA}
\author{C.~D.~Anson}\affiliation{Ohio State University, Columbus, Ohio 43210, USA}
\author{A.~Aparin}\affiliation{Joint Institute for Nuclear Research, Dubna, 141 980, Russia}
\author{D.~Arkhipkin}\affiliation{Brookhaven National Laboratory, Upton, New York 11973, USA}
% ...

\collaboration{STAR Collaboration}\noaffiliation


\begin{abstract}
We present a study of the average transverse momentum
($p_t$) fluctuations and $p_t$ correlations for charged particles
produced in Cu+Cu collisions at midrapidity for $\sqrt{s_{NN}} =$ 62.4 and 200
GeV. 
These results are compared with those published for Au+Au
collisions at the same energies, to explore the system size dependence. 
In addition to the collision energy and system size dependence, the $p_t$
correlation results have been studied as functions of the collision
centralities, the ranges in $p_t$, the pseudorapidity $\eta$, and the
azimuthal angle $\phi$.
The square root of the measured $p_t$ correlations when scaled by
mean $p_t$ is found to be independent of both colliding beam energy
and system size studied. Transport-based model calculations are
found to have a better quantitative agreement with the measurements
compared to models which incorporate only jetlike correlations.
\end{abstract}

\maketitle

\section{Introduction}
The study of event-by-event fluctuations and correlations is an important tool to understand
thermalization and phase transitions in heavy-ion collisions~\cite{heis,fluct,jetq2_5,fluct2,pmdfluc_5,adcox_5}. 
Non-monotonic change in transverse momentum ($p_t$) correlations as a function of centrality and/or the incident energy
has been proposed 
as a possible signal of quark gluon plasma (QGP) formation~\cite{heis}. 
The QGP is believed to be formed at the early stage of high energy heavy-ion collisions when the
system is hot and dense. As time passes, the system dilutes, cools down and hadronizes. 
Fluctuations are supposed to be sensitive to the dynamics of the
system, especially at the 
QGP to hadron gas transition~\cite{fluct2,adcox_5}. 
Alternatively, analyses at the Relativistic Heavy Ion Collider (RHIC)
based on $p_t$ auto-correlations (the inversion of the mean transverse momentum 
fluctuations in various pseudorapidity and azimuthal angle
difference regions of the produced particles) 
indicate that the basic correlation mechanism could be dominated by the process of parton 
fragmentation~\cite{parton}. 
Thus fluctuation measurements are proposed to be an important tool in understanding 
nucleus-nucleus collisions~\cite{heis,adcox_5,steph_5,steph2_5,steph3_5,steph4_5,dis_5,pmdfluc_5}.
In addition, under the assumption that thermodynamics is applicable to heavy-ion collisions, fluctuations 
in various observables could be related to thermodynamic properties of the
matter formed. 
For example, the event-by-event $\langle p_{t} \rangle$ 
could be related to temperature fluctuations~\cite{stodolsky_5,shuryak2_5,rajagopal_5,berdi_5,voloshin_5}.

The study of event-by-event fluctuations of various quantities 
such as eventwise mean transverse momentum 
($\langle p_{t} \rangle$), charged track multiplicity, and conserved quantities such
as net-baryon and/or net-charge number
are considered to be some of the main probes in the search for the critical point in the 
QCD phase diagram~\cite{voloshin_5,bass_5,bass2_5,jeon_5,asakawa_5,lin_5,heis2_5,shuryak_5,pruneau_5,gavin_5}. 
One expects enhanced fluctuations in the above observables when the
system passes through the vicinity of the critical point. 
Recent results from the 
CERN Super Proton Synchrotron 
(SPS)
experiments show
that the energy dependence of transverse momentum fluctuations does not show the increase expected for freeze-out 
near the critical point of QCD~\cite{na49_pt}. However, when these
fluctuations are studied as a function of the system size (colliding
C+C, Si+Si, Pb+Pb), enhanced fluctuations are observed in smaller colliding
systems~\cite{na49}.
The results from the RHIC beam energy scan (BES)~\cite{bes} data for the above
observables will provide further insights.

The results presented here are from Cu+Cu collisions at 
$\sqrt{s_{NN}} =$~62.4 and 200 GeV using the
solenoidal tracker At RHIC (STAR) and
are compared with the published results from
Au+Au collisions at the same energies~\cite{prc_200_5}.
This paper describes a systematic study of the system size dependence of
the transverse momentum correlations observed at RHIC.

The paper is organized as follows. 
The STAR detector, the data set, and the centrality
selection used in the analysis, are discussed in Sec. II.
In Sec. III, we discuss
$\langle p_{t} \rangle$ fluctuations extracted
from the $\langle p_{t} \rangle$ distributions, which are compared
with mixed events and gamma distributions. 
Dynamical fluctuations are extracted
and presented for Au+Au and Cu+Cu collisions at $\sqrt{s_{NN}}=$~62.4
and 200 GeV. 
The $p_{t}$ correlations and the dynamical correlations when scaled by $\langle
N_{\mathrm {part}} \rangle$ and $\langle p_{t} \rangle$ are discussed
in Sec. IV,
to understand the centrality and energy dependence. 
Experimental data is also compared to various model calculations. 
Finally, a detailed study of $\eta$, $\phi$ and $p_{t}$ dependence of
the correlations, is presented. 
The systematic errors
associated with the analysis are discussed in Sec. V.
We conclude with a summary in Sec. VI.

\section{Experiment and Data Analysis}
The Cu+Cu data were taken using the STAR detector with a minimum bias trigger. 
For the data taken at $\sqrt{s_{NN}}=$ 200 GeV this was done by
requiring a coincidence from the two zero degree calorimeters 
(ZDCs). For the $\sqrt{s_{NN}}=$ 62.4 GeV data, the ZDC is less efficient,
so a beam beam counter (BBC) coincidence was also required. 
More details about the trigger detectors 
can be obtained from Ref.~\cite{trig_5}.

The main detector used in this analysis is the time projection chamber
(TPC)~\cite{star_nim}, which is the primary tracking device in STAR.
The TPC is 4.2 m
 long and 4~m in diameter and its acceptance spans about $\pm$1.0 units of
 pseudorapidity and full azimuthal coverage. 
The sensitive volume of the TPC contains P10 gas (10\% methane, 90\% argon) regulated at 2 mbar above atmospheric
pressure. The TPC data are used to determine particle trajectories,
momenta, and particle-type through ionization energy loss ($dE/dx$)~\cite{bichsel}. 

The primary vertex of events used in this analysis is required to be within $\pm$ 30 cm of the geometric center of
the TPC along the beam axis.
This selection process provides about 7.5 $\times$ $10^6$ and 
15 $\times$ $10^6$ minimum bias events for Cu+Cu collisions at $\sqrt{s_{NN}} =$ 62.4 and 200 GeV,
respectively.

The collision centralities are represented by the fractions of events
in the minimum bias inelastic 
cross section distribution
in a collision. In data, the collision centrality is determined by using the 
uncorrected charged track multiplicity ($N_{\rm{ch}}^{\rm{TPC}}$), measured in the TPC 
within $|\eta| <$ 0.5. 
The various centrality 
bins are calculated as a fraction of this multiplicity distribution starting at 
the highest multiplicities.
The centrality classes for Cu+Cu collisions at $\sqrt{s_{NN}} =$ 62.4 and 200 GeV
are 0\%--10\% (most central), 10\%--20\%, 20\%--30\%, 30\%--40\%, 40\%--50\% and 50\%--60\%
(most peripheral). 
Each centrality bin is associated with an average number of participating nucleons
($\langle N_{\rm{part}} \rangle$)
obtained using Glauber Monte Carlo 
simulations~\cite{mc_5} employing the Woods-Saxon distribution for the nucleons inside the
Cu nucleus. 
The systematic uncertainties include those determined by varying the
Woods-Saxon parameters, those associated with the nucleon-nucleon
cross sections, those related to the functional representation of the
multiplicity distribution, and those associated with the 
determination of the total Cu+Cu cross section.
Table~\ref{tabcent62_5}  lists the 
$N_{\rm{ch}}^{\rm{TPC}}$ and $\langle N_{\rm{part}} \rangle$ 
values for each centrality in Cu+Cu collisions at 
$\sqrt{s_{NN}} =$ 62.4 and 200 GeV. 
Corresponding values for Au+Au collisions can be found in Ref.~\cite{mc_5}.

In order to have uniform detector performance, a pseudorapidity cut of
$|\eta| <$ 1.0 is 
applied to tracks. To avoid the admixture of tracks from a secondary vertex, a requirement is placed on the
distance of closest approach (DCA) between each track and the event vertex. 
The charged particle tracks are required to have originated within 1 cm of the measured 
event vertex.
The multiple counting of split tracks (more than one track
reconstructed from the original single track) is avoided by applying a condition on the number
of track fit points ($N_{\rm Fit}$) used in the reconstruction of the
track. 
Each included track is required to have a minimum
number of 21 TPC points along the track.
The transverse momentum range selected for the analysis is 0.15--2.0
GeV/$c$. 

The errors shown in the figures have the statistical and systematic
errors added in quadrature unless otherwise stated. The statistical errors are small and are
within symbol sizes. The final systematic errors are
obtained as quadrature sums of systematic errors from different
sources as discussed in Sec. V.

\section{$\langle p_{t} \rangle$ Fluctuations}
The $p_{t}$ fluctuations in high energy collisions can be measured using the 
distribution of the event-wise mean transverse momentum defined
as 
\begin{equation}
\label{mpteq_5}
\langle p_{t} \rangle= \frac{1}{N} \sum_{i=1}^{N}p_{t,i},
\end{equation}
where $N$ is the multiplicity of accepted tracks 
from the primary vertex in a given event and $p_{t,i}$ is the
transverse momentum 
of the $i{\rm{th}}$ track. The 
mean-$p_{t}$ distribution is compared to the corresponding distribution
obtained for mixed events~\cite{prc_200_5}. 
Mixed events are constructed by randomly
selecting one track from an event chosen from the events in the same
centrality and same event vertex bin. The mixed events are created
with the same multiplicity distribution as that of the real events.
\begin{table}
\caption{\label{tabcent62_5}
The $N_{\rm {ch}}^{\rm{TPC}}$ values and average number of participating nucleons ($\langle N_{\rm{part}} \rangle$) 
for various collision centralities in Cu+Cu collisions at
$\sqrt{s_{NN}}$ = 62.4 and 200 GeV.}
\begin{center}
\begin{tabular}{c|c|c|c|c}
\hline
\% cross &\multicolumn{2}{l|}{Cu+Cu 62.4
   GeV}&\multicolumn{2}{l}{Cu+Cu 200 GeV}\\  [0.1cm]
\cline{2-5} \\ [-0.3cm]
 section&$N_{\rm {ch}}^{\rm{TPC}}$&$\langle N_{\rm {part}} \rangle
 $&$N_{\rm {ch}}^{\rm{TPC}}$ &$\langle N_{\rm {part}} \rangle $\\ [0.2cm]
\hline
0-10    & $ >101$   & $96.4^{+1.1}_{-2.6}$   & $ >139$   & $99.0^{+1.5}_{-1.2}$     \\ [0.1cm]
10-20  & $71-100$   &$72.2^{+0.6}_{-1.9}$  & $98-138$   &$74.6^{+1.3}_{-1.0}$   \\ [0.1cm]
20-30  & $49-70$   & $51.8^{+0.5}_{-1.2}$   & $67-97$   & $53.7^{+1.0}_{-0.7}$  \\[0.1cm]
30-40  & $33-48$   & $36.2^{+0.4}_{-0.8}$   & $46-66$   & $37.8^{+0.7}_{-0.5}$   \\[0.1cm]
40-50  & $22-32$   & $24.9^{+0.4}_{-0.6}$  & $30-45$   & $26.2^{+0.5}_{-0.4}$     \\[0.1cm]
50-60  & $14-21$   & $16.3^{+0.4}_{-0.3}$   & $19-29$   & $17.2^{+0.4}_{-0.2}$     \\[0.1cm]
\hline
\end{tabular}
\end{center}
\end{table}
\begin{figure}
\begin{center}
\includegraphics[scale=0.4]{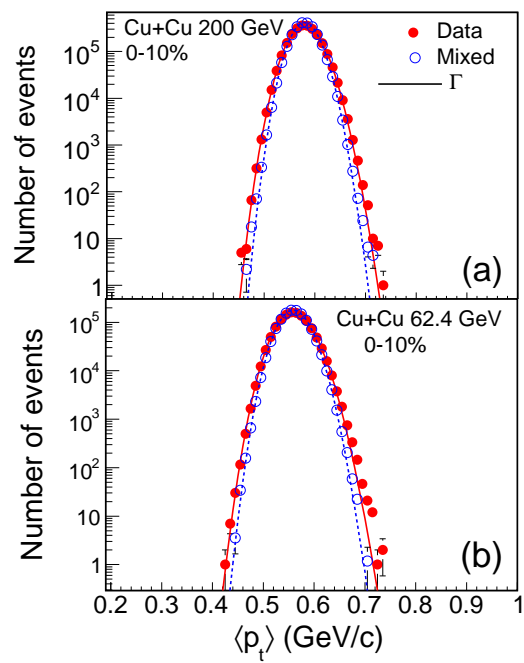}
\caption{ (Color online) Event-by-event $\langle p_{t} \rangle$ distributions for
data and mixed events in central
Cu+Cu collisions at (a) $\sqrt{s_{NN}} =$ 200 and (b) 62.4 GeV. The
curves (solid for data and dotted for mixed events)
represent the $\Gamma$ 
distributions.
The errors shown are statistical.
}
\label{figmpt62_5}
\end{center}
\end{figure}

\begin{table}
\caption{\label{tabmeansigma62_5}
Gamma distribution parameters for event-by-event $\langle p_{t}
\rangle$  distributions for data and mixed events in central
(0\%-10\%) Cu+Cu collisions 
at $\sqrt{s_{NN}} =$ 62.4 and 200 GeV. 
}
\begin{center}
\begin{tabular}{c|c|c|c|c}
\hline
 Collision         & $\alpha$ & $\beta$ ($\times 10^{-3}$           &  $\mu$   & $\sigma$  \\ 
 type (AA)        &                 &  GeV/$c$)    &  (GeV/$c$)
 &  (GeV/$c$)           \\ 
\hline
Cu 200 (data)             &    476   &  1.22      &0.5805 & 0.02660  \\
Cu 200 (mixed)   &    634   &  0.92     &0.5807 & 0.02310   \\
Cu 62.4 (data)            &     358   &  1.56      &0.5603 & 0.02960 \\
Cu 62.4 (mixed)  &     457   &  1.23      &0.5602 & 0.02621 \\
\hline
\end{tabular}
\end{center}
\end{table}
Figure~\ref{figmpt62_5} shows the event-by-event mean-$p_{t}$
distributions for 0\%-10\% Cu+Cu collisions at
(a) $\sqrt{s_{NN}} =$ 200 and (b) 62.4 GeV.
The solid symbols represent
the $\langle p_{t} \rangle$ distributions for data; the open symbols represent
$\langle p_{t} \rangle$ distributions for mixed events. 
The distributions are similar for other centralities.
The mixed events
provide a reference measure of statistical fluctuations in the data. 
Any fluctuations
observed in data beyond these statistical fluctuations are referred to as
non-statistical or dynamical fluctuations in this paper.
For both data and mixed events, 
while going from central to peripheral collisions, 
the mean ($\mu$) of the distributions decreases 
whereas
the
standard deviation ($\sigma$) increases.
Moreover, it is seen that the $\langle p_{t} \rangle$ distributions
for data are wider than those for mixed 
events, suggesting the presence of non-statistical fluctuations in
Cu+Cu data for both 62.4 and 200 GeV collisions. 

The curves in Fig.~\ref{figmpt62_5} represent 
the gamma ($\Gamma$) distributions
for data (solid lines) and mixed events (dotted lines). The $\Gamma$
distribution~\cite{fluct2,tannen_5} is given by
\begin{equation}
\label{gammaeq_5}
f(x) = \frac{x^{\alpha - 1}  e^{-x/\beta}}{\Gamma(\alpha)\beta^{\alpha}},
\end{equation}
where $x$ represents the $\langle p_{t} \rangle$;
$\alpha=\mu^{2}/\sigma^{2}$ and $\beta=\sigma^{2}/\mu$.

Tannenbaum~\cite{tannen_5} argues that the 
$\Gamma$ distribution 
is one of the standard representations of the inclusive single particle
$p_{t}$ distribution. 
Tannenbaum~\cite{tannen_5}
also suggests that the quantity $\alpha/\langle N_{\rm{ch}}
\rangle$, should be $\sim$ 2, and the quantity $\beta \times\langle N_{\rm{ch}} \rangle$ 
representing the inverse slope parameter may be referred to as the temperature 
of the system. 
 Here $\langle N_{\rm{ch}} \rangle$ is the average charged particle 
multiplicity in a given centrality bin. 
It is found that $\alpha/\langle N_{\rm{ch}} \rangle$ for Cu+Cu 0\%--10\%
central collisions 
is 2.04 at 200 GeV, and is 2.18 at 62.4 GeV.
The respective $\beta \times\langle N_{\rm{ch}} \rangle$ values are
0.284 GeV/$c$ and 0.256 GeV/$c$. 
The $\alpha/\langle N_{\rm{ch}} \rangle$ and $\beta \times\langle
N_{\rm{ch}} \rangle$  for 0\%--5\% central Au+Au collisions at 200 GeV were
found to be 1.93 and 0.299 GeV/$c$~\cite{prc_200_5}.
Table~\ref{tabmeansigma62_5}  lists
gamma distribution parameters for event-by-event $\langle p_{t}
\rangle$  distributions for data and mixed events in central 
(0\%--10\%) Cu+Cu
collisions at $\sqrt{s_{NN}} =$ 62.4 and 200 GeV. 
The values for Au+Au collisions can be found in
Ref. ~\cite{prc_200_5}.
For Cu+Cu collisions at 200 GeV, $\alpha/\langle N_{\rm{ch}}
\rangle$ varies from 2.04 to 2.11 from central to peripheral
collisions. However, for Cu+Cu at 62.4 GeV, 
$\alpha/\langle N_{\rm{ch}} \rangle$
varies from 
2.18 to 2.27 from central to peripheral collisions. The errors on
values of $\alpha$ and $\beta$ are of the order of 13--18\% and
9-12\%, respectively,
for Cu+Cu collisions. 

\begin{figure}[htb]
\begin{center}
\includegraphics[scale=0.4]{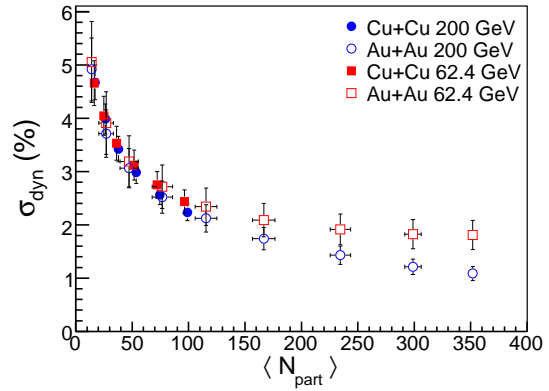}
\caption{
(Color online) Comparison of dynamical $\langle p_{t} \rangle$  fluctuations in Au+Au and Cu+Cu collisions 
at  $\sqrt{s_{NN}} =$ 62.4 and 200 GeV as a function of the number of participating 
nucleons.
}
\label{figsigdyn_5}
\end{center}
\end{figure}

The non-statistical or dynamical fluctuations in mean-$p_t$
are quantified using a variable $\sigma_{\rm{dyn}}$~\cite{sigmdyn_ref,k2pi_star}
defined as
\begin{equation}
\label{eqdynfluc_5}
\sigma_{\rm{dyn}} = \sqrt{ \left(\frac{\sigma_{\rm{data}}}{\mu_{\rm{data}}} \right)^{2} -
\left(\frac{\sigma_{\rm{mix}}}{\mu_{\rm{mix}}} \right)^{2}},
\end{equation}
where $\mu_{\rm{data}}$ and $\mu_{\rm{mix}}$ are the means of the event-by-event $\langle p_{t} \rangle $ 
distributions for data and mixed events, respectively. Similarly, $\sigma_{\rm{data}}$ and 
$\sigma_{\rm{mix}}$ are respectively the standard deviations of $\langle p_{t} \rangle $ 
distributions for data and mixed events.
Figure~\ref{figsigdyn_5} shows the dynamical fluctuations ($\sigma_{\rm{dyn}}$) in mean-$p_{t}$ plotted as
a function of $\langle N_{\rm{part}} \rangle $. 
The results are shown for Cu+Cu collisions at
$\sqrt{s_{NN}} =$ 62.4 and 200 GeV, and are compared with the results from Au+Au collisions at
$\sqrt{s_{NN}} =$ 62.4 and 200 GeV.
The dynamical  $\langle p_{t} \rangle$  fluctuations are similar in
Au+Au and Cu+Cu collisions
at similar values of $\langle N_{\rm{part}} \rangle $.
The fluctuations decrease as $\langle N_{\rm{part}} \rangle $
increases. 
The dynamical fluctuations are also independent of the collision
energy and are found to vary
from $\sim$ 2\% to $\sim$ 5\% 
for $\langle N_{\rm{part}} \rangle $ less than $\sim$120.
For $\langle N_{\rm{part}} \rangle $ greater than $\sim$150,
the dynamical fluctuations are smaller and vary from 1\% to 2.5\%.

\section{$p_{t}$ Correlations}
Non-statistical or dynamical fluctuations can also be analyzed by using 
two-particle transverse momentum correlations~\cite{prc_200_5}.
The two-particle $p_{t}$ correlations are studied using the 
following equation~\cite{prc_200_5}: 
\begin{equation}
\label{corr}
\langle \Delta p_{t,i}~\Delta p_{t,j} \rangle=\frac{1}{N_{\rm {event}}}
\sum_{k=1}^{N_{\rm {event}}}{\frac{C_{k}}{N_{k}(N_{k}-1)}},
\end{equation}
where $C_{k}$ is the two-particle transverse momentum covariance for
the $k{\rm{th}}$ event,
\begin{equation}
\label{cov_5}
C_{k}=\sum\limits_{i=1}^{N_{k}} {\sum\limits_{j=1,i\ne j}^{N_{k}}
{{\left( {p_{t,i}-\left\langle {\left\langle {p_t} \right\rangle }
\right\rangle } \right)\left( {p_{t,j}-\left\langle
{\left\langle {p_t} \right\rangle } \right\rangle } \right)} }},
\end{equation}
where $p_{t,i}$ is the transverse
momentum of the $i{\rm{th}}$ track in the $k{\rm{th}}$ event, $N_k$ is the number of
tracks in the $k{\rm{th}}$ event, and $N_{\rm {event}}$ is the number of events. The overall event average transverse 
momentum ($\langle\langle p_{t} \rangle \rangle$) is given by 
\begin{equation}
\label{meanmpt_5}
\langle\langle p_{t} \rangle\rangle = \left({\sum\limits_{k=1}^{N_{\rm{event}}} {\langle p_{t} \rangle}_{k}}\right)
\slash N_{\rm{event}},
\end{equation}
where ${\langle p_{t} \rangle}_{k}$ is the average transverse 
momentum in the $k{\rm{th}}$ event. 
It may be noted that, for mixed events, there will be no dynamical
fluctuations and the $p_{t}$ correlations will be zero.
Equation~(\ref{corr}) is used to obtain the $p_{t}$ correlations in 
Cu+Cu collisions at $\sqrt{s_{NN}} =$ 62.4 and 200 GeV.
These 
results are compared with the published results from Au+Au collisions at similar 
energies~\cite{prc_200_5}
to investigate the system-size and collision energy dependence of the $p_{t}$
correlations in heavy-ion collisions at RHIC.

The $p_t$ correlation values may be influenced by the dependence of the correlations 
    on the size of the centrality bin due to variation of $\langle \langle {p_{t}} \rangle \rangle$ with centrality.
    This dependence is removed by calculating $\langle \langle {p_{t}} \rangle \rangle$ as a function of $\langle N_{\rm{ch}} \rangle$, which is the multiplicity 
    of charged tracks used to define the centrality. This multiplicity dependence of $\langle \langle {p_{t}} \rangle \rangle$ is fitted
    with a suitable polynomial in $\langle N_{\rm{ch}} \rangle$ and used in Eq.~(\ref{cov_5}) for $\langle \langle {p_{t}} \rangle \rangle$.
All results presented in this paper have been corrected for
this effect.

\begin{figure}[htb]
\begin{center}
\includegraphics[scale=0.6]{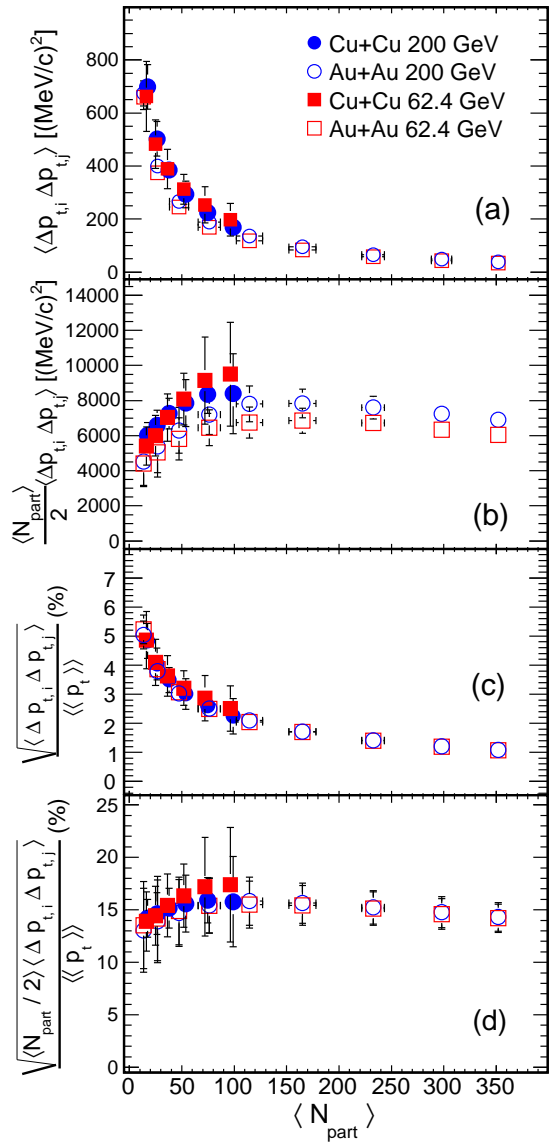}
\caption{(Color online) (a) $p_{t}$ correlations, (b) $p_{t}$ correlations
  multiplied by $\langle N_{\rm{part}} \rangle/2$, (c) square root of
  $p_{t}$ correlations scaled by $\langle \langle p_{t} \rangle
  \rangle$, and (d) square root of $p_{t}$ correlations multiplied by
  $\langle N_{\rm{part}}\rangle/2$ and scaled by $\langle \langle
  p_{t} \rangle \rangle$, plotted vs. $\langle N_{\rm{part}}
  \rangle$. Results are compared between Cu+Cu and Au+Au collisions at
  $\sqrt{s_{NN}} =$ 62.4 and 200 GeV.  Au+Au data have been taken from Ref.~\cite{prc_200_5}.
}
\label{figcorrvsnp_5}
\end{center}
\end{figure}

Figure~\ref{figcorrvsnp_5} (a) shows the $p_{t}$ correlations plotted
as a function of 
$\langle N_{\rm{part}} \rangle$ for Cu+Cu and Au+Au collisions at  $\sqrt{s_{NN}} =$ 62.4 and 200 GeV. 
The $p_{t}$ correlations 
decrease with increasing $\langle N_{\rm{part}} \rangle$ for Au+Au and
Cu+Cu at both energies. 
The decrease in correlations with increasing participating nucleons
could 
suggest that
correlations are dominated by pairs of particles that
originate from the same nucleon-nucleon collision, and they get diluted when the  
number of participating nucleons increases~\cite{prc_200_5}.

\subsection{Scaled $p_{t}$ Correlations} 
The decrease in $p_t$ correlations with increasing $\langle N_{\rm{part}}
\rangle$ seen in Fig.~\ref{figcorrvsnp_5}~(a) 
may be related to a system volume dependence characterized by $\langle
N_{\rm{part}} \rangle$.  
This volume dependence is removed by  
multiplying the $p_t$ correlations by $\langle N_{\rm{part}} \rangle/2$ as
shown in Fig.~\ref{figcorrvsnp_5} (b).
The results are shown for Cu+Cu and 
Au+Au collisions at $\sqrt{s_{NN}} =$ 62.4 and 200 GeV. It is observed that this measure 
of $p_t$ correlations increases quickly with increasing  $\langle N_{\rm{part}} \rangle$ for both Cu+Cu and
Au+Au collisions and saturates for central Au+Au collisions. 
The saturation of this quantity might indicate effects
such as the onset of thermalization~\cite{gavin_5},  
the onset of jet quenching~\cite{steph4_5,jetq2_5}, 
or the saturation of transverse flow in central collisions~\cite{tflow_5}.
It seems that, for Cu+Cu collisions, this quantity is larger than for
Au+Au collisions which might indicate more correlations for the
smaller systems. 
However, the size of the errors in the current analysis does not allow a conclusive statement. 

The correlation measure $\langle \Delta p_{t,i}~\Delta p_{t,j} \rangle$ may change
due to changes in $\langle \langle p_{t} \rangle \rangle$ with incident energy and/or
collision centrality. To address these changes, 
the square roots of the
measured correlations are scaled by $\langle \langle p_{t} \rangle \rangle$. 
Figure~\ref{figcorrvsnp_5} (c) shows the corresponding quantity
$\sqrt{\langle \Delta p_{t,i}~\Delta p_{t,j} \rangle} / \langle \langle p_{t} \rangle \rangle$
plotted as a function of collision centrality for Cu+Cu and Au+Au collisions at 
$\sqrt{s_{NN}} =$ 62.4 and 200 GeV. 
It is observed that the correlation scaled by $\langle \langle p_{t} \rangle \rangle$
is independent of collision system size and energy, but decreases with increasing  $\langle N_{\rm{part}} \rangle$.
The combined effect of multiplying $p_{t}$ correlations by  $\langle
N_{\rm{part}}\rangle/2$ and scaling with $\langle \langle p_{t}
\rangle \rangle$ is shown in Fig.~\ref{figcorrvsnp_5} (d). It seems that
this quantity $\sqrt{\langle N_{\rm{part}}/2\rangle \langle \Delta
  p_{t,i}~\Delta p_{t,j} \rangle} / \langle \langle p_{t} \rangle
\rangle$ increases with $\langle N_{\rm{part}}\rangle$ and shows
saturation for central Au+Au collisions, but is independent of
collision system and energy.

\subsection{Model Comparisons}
It is interesting to compare the above 
results with theoretical model calculations to understand the physical mechanism behind these
measurements. 
The comparison is made with some widely used models in
heavy-ion collisions such as ultrarelativistic quantum molecular
dynamics (URQMD)~\cite{urqmd}, a multiphase transport model (AMPT) (default and string-melting)~\cite{ampt},
and the heavy-ion jet interaction generator (HIJING) (with jet quenching
switched off and on)~\cite{hijing_5}. 
The model results are obtained using UrQMD version 2.3, AMPT version
1.11 for default, and version 2.11 for AMPT string-melting.

HIJING is a
perturbative QCD-inspired model that
produces multiple minijet partons; these later get transformed into
string configurations and then fragment to hadrons. The fragmentation
is based on the Lund jet fragmentation model~\cite{lund_model}. A parametrized
parton-distribution function inside a nucleus is used to take into
account parton shadowing. 

AMPT uses the same initial conditions as in HIJING. However, the
minijet partons are made to undergo scattering before they are allowed
to fragment into hadrons. The string-melting (SM) version of the AMPT
model (labeled here as AMPT Melting) is based on the idea
that for energy
densities beyond a critical value of 1~GeV/$\rm{fm}^3$, 
the system should be de-confined and strings (or hadrons) decomposed into their partonic components.
This is done by converting
the mesons to a quark-antiquark pair, baryons to three quarks, and
so on. The scattering of the quarks is based on a parton cascade.
Once the interactions stop, the partons then hadronize through
the mechanism of parton coalescence. 
\begin{figure}
\begin{center}
\includegraphics[scale=0.48]{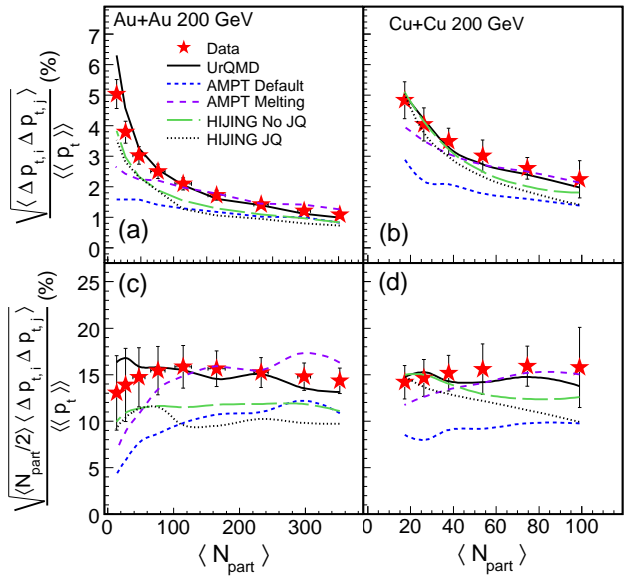}
\caption{(Color online) Comparison of scaled $p_{t}$ correlations between
 data and models for Au+Au [panels (a) and (c)] and Cu+Cu [panels (b)
 and (d)] collisions at 200 GeV. 
Au+Au data has been taken from Ref.~\cite{prc_200_5}. The
curves represent different model calculations.
}
\label{corr_w_mod}
\end{center}
\end{figure}
\begin{figure}
\begin{center}
\includegraphics[scale=0.4]{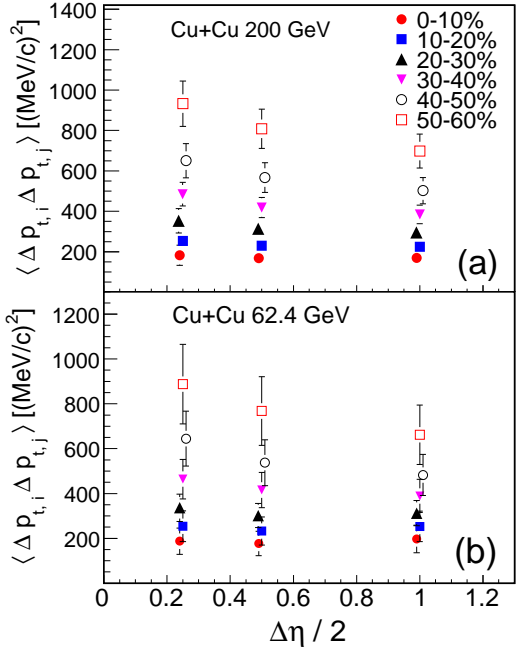}
\caption{(Color online) $p_{t}$ correlations for varying rapidity
  acceptance ($|\eta|<$0.25, 0.5, and 1.0) for Cu+Cu collisions at
  $\sqrt{s_{NN}}=$ 200 GeV [panel (a)] and
 62.4 GeV [panel (b)]. 
}
\label{corr_eta}
\end{center}
\end{figure}
The URQMD model is based on a microscopic transport theory where the 
phase-space description of the reactions is important. It allows for
the propagation of all hadrons on classical trajectories in
combination with stochastic binary scattering, color string formation,
and resonance decay. 

Figure~\ref{corr_w_mod} shows the comparison of $p_{t}$ correlations
between data [(a) and (c) for Au+Au 200 GeV, and (b) and (d) for Cu+Cu
200 GeV] and the 
models described above.
The 
transport-based URQMD model calculations are observed to have a better
quantitative agreement with the measurements compared to models which 
incorporate only jetlike correlations as in HIJING. HIJING gives
similar dependence on $\langle N_{\rm{part}} \rangle$ but under-predicts the magnitude. Inclusion
of the jet-quenching effect in HIJING leads to a smaller value of the
correlations in central collisions. 
AMPT model calculation
 incorporating coalescence as a mechanism of particle
production also compares well with data for central collisions. However
the default version of this model which incorporates additional initial and final state 
scattering relative to HIJING yields smaller values of correlations.

\subsection{$\eta$ and $\phi$ Dependence}
The $\eta$ and $\phi$ dependences of $p_{t}$ 
correlations are also studied. 
Figure~\ref{corr_eta} shows the $p_{t}$ correlations plotted as 
a function of increasing rapidity acceptance for Cu+Cu collisions at
(a) $\sqrt{s_{NN}} =$ 200 and (b) 62.4 GeV.
The data points for centralities 0\%-10\%, 20\%-30\%, and 40\%-50\% are shifted
by 0.01 in $\Delta\eta$/2 for clarity. The correlations
are almost independent of the $\Delta\eta$ window for
the most central collisions. For peripheral
collisions, the correlations show a slight
rapidity dependence with maximum value for
$-0.25 < \eta  < 0.25$.

\begin{figure}
\begin{center}
 \includegraphics[scale=0.46]{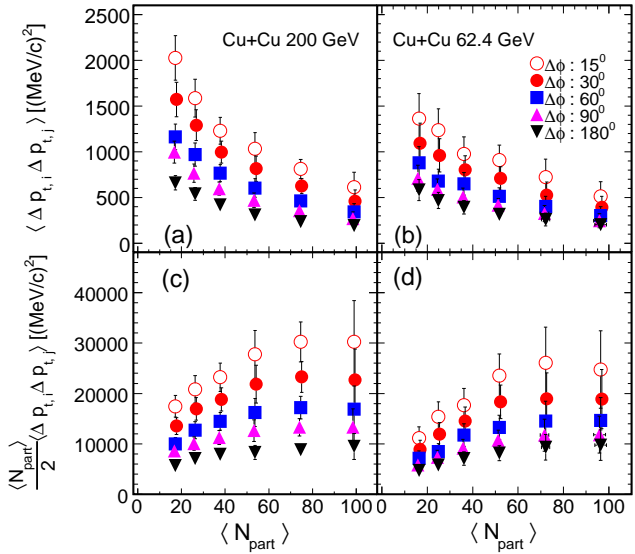}
\caption{(Color online) $p_{t}$ correlations for varying azimuthal
  acceptance
for Cu+Cu collisions at $\sqrt{s_{NN}}=$ 200 GeV [panels (a) and (c)] and
 62.4 GeV [panels (b) and (d)]. 
}
\label{corr_phi}
\end{center}
\end{figure}
\begin{figure}
\begin{center}
\includegraphics[scale=0.4]{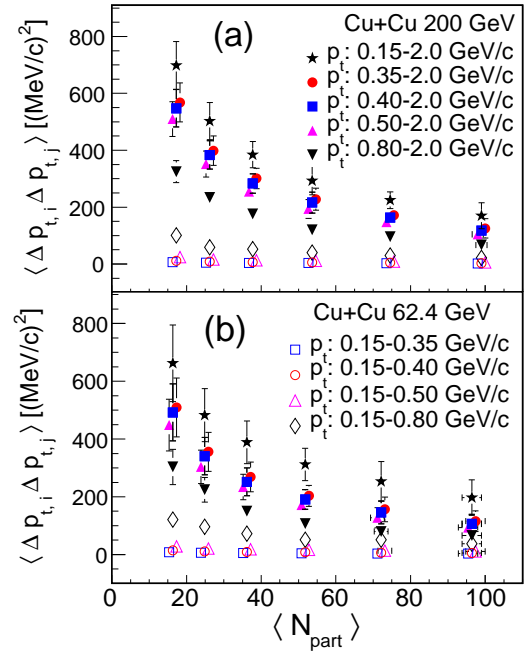}
\caption{(Color online) 
$p_t$ correlations plotted as a function of $\langle N_{\rm{part}}
\rangle$  for different $p_{t}$ ranges  
in Cu+Cu collisions at (a) 200 GeV  and (b) 62.4 GeV. 
}
\label{corr_pt}
\end{center}
\end{figure}

Figure~\ref{corr_phi} shows the $p_{t}$ correlations for varying 
azimuthal angle windows 
for Cu+Cu collisions at
(a) $\sqrt{s_{NN}} =$ 200 and (b) 62.4 GeV.
The data points for $\Delta\phi$ windows: $30^{0}$ and $90^{0}$
are shifted by 0.5 units in $\langle N_{\rm{part}} \rangle$ for clarity. 
The $p_t$ correlations are maximum for $\Delta\phi = 15^{0} $
(among the cases studied) and decrease with increasing
$\langle N_{\rm{part}} \rangle$
 for a given $\Delta\phi$ window, 
as expected.
The $p_t$ correlations multiplied by $\langle N_{\rm{part}} \rangle$/2
[Fig.~\ref{corr_phi} (c) for Cu+Cu 200 GeV and Fig.~\ref{corr_phi} (d) for Cu+Cu
62.4 GeV] seem to increase and then saturate with increasing  $\langle N_{\rm{part}} \rangle$. 

\subsection{$p_{t}$ Dependence}
Figure~\ref{corr_pt} shows the correlations as a function of collision centrality for different
$p_{t}$ regions in Cu+Cu collisions at (a) $\sqrt{s_{NN}} =$ 200 and
(b) 62.4 GeV.
The different $p_{t}$ ranges used are shown. 
These $p_t$ ranges are chosen to demonstrate the dependence of the correlations among tracks sets at lower $p_t$, at higher $p_t$, and in a set where all available $p_t$ values are included.
The data points for $p_{t}$ ranges (in GeV/$c$): 0.15-0.5, 0.5-2.0,
0.15-0.35, and 0.35-2.0, are shifted by one unit in $\langle N_{\rm
  part} \rangle$. 

The $p_t$
correlation is maximum (minimum) for charged particles
whose $p_{t}$ is in 0.15--2.0 GeV/$c$ (0.15--0.35 GeV/$c$). 
The $p_{t}$ correlation values are small and fairly independent of $p_{t}$ if
a lower $p_{t}$ bound for the particles studied is fixed at 0.15 GeV/$c$ and the
upper $p_{t}$ bound is progressively increased up to 0.50 GeV/$c$. When the
analysis is carried out by keeping the higher $p_{t}$ bound fixed at 2.0
GeV/$c$ and subsequently decreasing the lower $p_{t}$ bound to 0.15 GeV/$c$,
the correlation values are found to increase. 

\begin{figure}
\begin{center}
\includegraphics[scale=0.45]{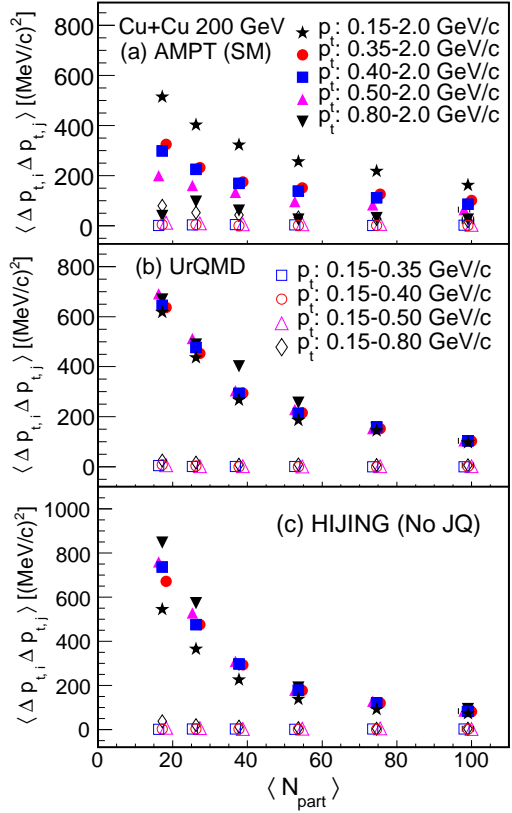}
\caption{(Color online) Correlations for varying $p_{t}$ ranges 
  for different model calculations in Cu+Cu 200 GeV: (a) AMPT (SM),
  (b) URQMD, and (c) Hijing (no JQ).
}
\label{corr_mod}
\end{center}
\end{figure}

Figure~\ref{corr_mod} shows the variation of  $p_{t}$ correlations as a
function of $\langle N_{\rm{part}} \rangle$ for different  $p_{t}$
windows as calculated using the (a) AMPT (string-melting),
(b) URQMD, and (c) HIJING (no jet quenching)  model calculations for Cu+Cu
collisions at 200 GeV. 
The AMPT calculations show $p_{t}$ correlations that are similar to
those observed in data for corresponding variations in the $p_t$
windows. 
The trend of the correlation values shown by both URQMD and HIJING is similar 
to what is seen in the data for the low-$p_t$ windows where the lower
bound is fixed at 0.15 GeV/$c$ and the upper bound is increased from
0.35 GeV/$c$ to 0.50 GeV/$c$. 
However, for URQMD, 
if the higher $p_t$ bound is fixed at 2.0 GeV/$c$ and the lower $p_t$ bound
is subsequently decreased to 0.15 GeV/$c$, $p_t$ correlation values
remain similar.
For the same case, 
HIJING
shows a decrease in $p_t$ correlation values when the lower $p_t$ bound is
decreased to 0.15 GeV/$c$.
This is just the opposite of what is observed
in data as seen in Fig.~\ref{corr_pt}.

\begin{figure}
\begin{center}
\includegraphics[scale=0.42]{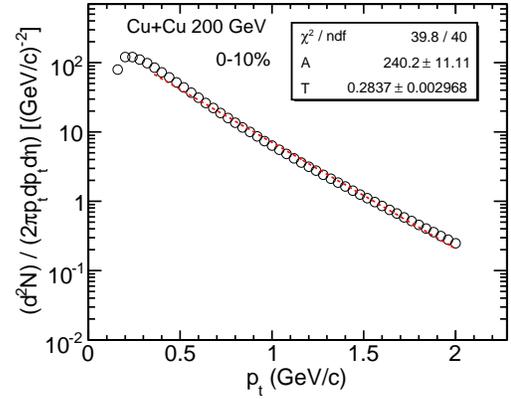}
\includegraphics[scale=0.45]{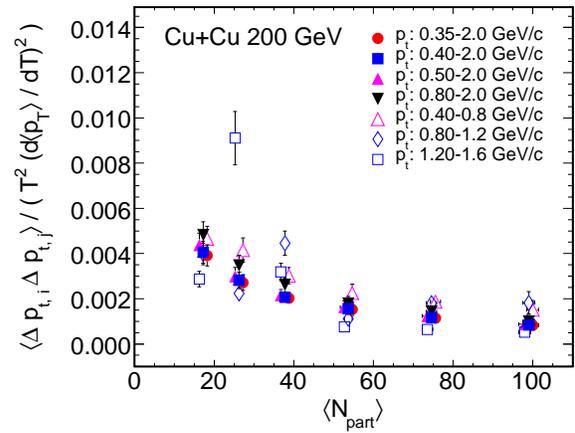}
\caption{(Color online) Top panel: Uncorrected inclusive charged particle $p_{t}$
  spectrum for 0\%--10\% collision centrality for Cu+Cu 200 GeV (open circles). 
 The distribution is a fit to the exponential function: $Ae^{-p_t/T}$
 (red dashed line).
 Errors are statistical. Bottom panel: Correlations scaled by $(d\langle p_t\rangle/dT)^2T^2$ vs. $\langle N_{\rm
  part} \rangle$ for different $p_t$ ranges in Cu+Cu 200 GeV. 
}
\label{corr_ptbin}
\end{center}
\end{figure}

Because correlations are calculated for different $p_{t}$ ranges,
the $p_{t}$ acceptance effect on the 
observed $p_t$ correlations, is examined. 
The correlation values in different $p_{t}$ ranges may depend on the
$p_{t}$ range size and on the 
fluctuations in the $p_t$ spectrum slope in that $p_{t}$
range. It is, therefore, important to see the effect of slope fluctuations
on the correlation values in different $p_{t}$ ranges.
The $p_{t}$ correlations can be formulated in terms of the 
fluctuations in the inverse slope parameter (effective temperature) by 
the following relation~\cite{voloshin_5}:
\begin{equation}
\label{ptbineq}
\langle \Delta p_{t}~\Delta p_{t} \rangle \approx \left[
  \frac{d\langle p_{t}\rangle }{dT} \right]^{2} \Delta T^{2},
\end{equation}
where $\Delta T^2$ describes the fluctuation in the inverse slope
parameter. The dependence $\langle p_t (T) \rangle$
can be obtained from the function
that describes the inclusive uncorrected $p_{t}$ spectrum for the desired $p_t$
range in the following manner. Figure~\ref{corr_ptbin} (top panel)
shows the measured inclusive
$p_{t}$ spectrum for 0\%--10\% collision centrality in Cu+Cu collisions
at 200 GeV. The dashed line represents the 
exponential fit of the form $F(p_t)= Ae^{-p_t/T}$ 
that is fit to these measurements for the 
$p_{t}$ range 0.35--2.0 GeV/$c$. 
The expressions for $\langle p_t \rangle$ and $(d\langle p_t
\rangle/dT)$ can be obtained using the following relation:
\begin{equation}
\label{mptfn1eq}
\langle p_{t} \rangle = \frac{\displaystyle\int_a^bp_{t}^{2}F(p_{t})dp_{t} }{\displaystyle\int_a^bp_{t}F(p_{t})dp_{t}},
\end{equation}
which gives,
\begin{equation}
\label{mptfn1eq2}
\langle p_{t} \rangle =  2T +  \frac{a^{2}  e^{-a/T} - b^{2} e^{-b/T} }{(a+T)e^{-a/T} - (b+T)e^{-b/T} }.
\end{equation}
Here, $a$ and $b$ are the lower and upper limits of a given $p_{t}$
range, respectively. 
The derivative of $\langle p_{t} \rangle$ with respect to $T$ is obtained as:
\begin{equation}
\label{dmptfn1eq}
\frac{d\langle p_{t} \rangle}{dT} =  2 -  \frac{Ae^{-(a+b)/T} + a^{2}e^{-2a/T} + b^{2}e^{-2b/T}}{[ (a+T)e^{-a/T} - (b+T)e^{-b/T} ]^{2}},
\end{equation}
where 
\begin{equation}
\label{dmptfn1eqA}
A =  \frac{ab(b-a)^{2}}{T^{2}} + \frac{(b^{2}-a^{2})(b-a)}{T}-(a^{2}+b^{2}).
\end{equation}
Using the lower and upper limits of a given $p_t$ range, and the corresponding $T$
from the spectrum fit for each $p_t$ range and collision
centrality, 
$(d\langle p_t\rangle/dT)^2$ are obtained 
for every $p_t$ range in each centrality.

Figure~\ref{corr_ptbin} (bottom panel) shows
the measured two-particle $p_t$ correlations scaled by $(d\langle
p_t\rangle/dT)^2T^2$ vs. $\langle N_{\rm
  part} \rangle$ for different $p_t$ ranges in Cu+Cu 200 GeV. 
The data points for $p_{t}$ ranges (in GeV/$c$): 0.4--0.8, 0.5--2.0,
1.2--1.6, and 0.35--2.0, are shifted by one unit in $\langle N_{\rm
  part} \rangle$ for clarity. 
The scaled $p_t$ correlations for different $p_t$ ranges
become similar and show little dependence on the collision
centrality. This study seems to suggest that the difference in the
$p_t$ correlations for different $p_t$ ranges may due to the fluctuations in slope of the
$p_t$ spectrum in those $p_t$ ranges.

\section{Systematic Uncertainties}

Systematic errors on the mean ($\mu$) and standard deviation ($\sigma$) in the 
$\langle p_{t} \rangle$ distributions (discussed in Sec. III), and $p_{t}$ correlations 
(discussed in Sec. IV) are mainly evaluated by varying the different
cuts used in the analysis,
re-doing the analysis using these changed cuts and determining the
resulting changes in the values of $\mu$, $\sigma$, and the $p_{t}$
correlations.
The difference is taken as the systematic error due to a particular
analysis cut.
The resulting systematic uncertainties, described below, are shown in
Tables~\ref{tabmpt62_5} and ~\ref{tabcorr62_5}
as a percentage of the result ($\mu$, $\sigma$, and the $p_{t}$
correlations) for various centralities for Cu+Cu collisions at both
62.4 and 200 GeV.

To study the effect of the z-vertex ($V_{z}$) cut, 
the $V_{z}$ acceptance is increased to $\pm$ 50 cm from the default value of $\pm$30 cm.
No change in $\mu$ or $\sigma$ or in the $p_{t}$ correlations is observed
when using the wider $V_{z}$.

The effect of the cuts used to suppress background tracks is studied
by changing the DCA cut from the default, DCA $<$ 1 cm, to DCA $<$ 1.5
cm and separately, changing the required number of fit points along
the track, $N_{\rm{Fit}}$, from the default $N_{\rm Fit}$ $>$ 20 to
$N_{\rm Fit}$ $>$ 15. The resulting systematic errors due to these
changes are listed in 
Tables~\ref{tabmpt62_5} and ~\ref{tabcorr62_5}
in the columns labeled ``DCA'' and ``$N_{\rm Fit}$''.

The effect of the size of the centrality bin on the $p_{t}$
correlations is addressed by fitting $\langle\langle p_{t}
\rangle\rangle$ as a function of $\langle N_{\rm{ch}} \rangle$ (see
Sec. IV).
To determine the systematic uncertainty associated with this process,
different polynomial functions are used to fit $\langle\langle p_{t}
\rangle\rangle$ vs. $\langle N_{\rm{ch}} \rangle$. The systematic
errors associated with this correction are listed in
Table~\ref{tabcorr62_5} 
in the columns labeled ``Poly.''.

The systematic uncertainty on the $p_t$ correlations that
may be associated with the application of the low-$p_t$ 
cut is estimated by removing this $p_t$ cut in the HIJING~\cite{hijing_5} model
calculations. The estimated systematic errors are shown in
Table~\ref{tabcorr62_5}
in the columns labeled ``Low $p_t$''.

\begin{table}
\caption{\label{tabmpt62_5}
Systematic errors on $\mu$ and $\sigma$ in event-wise $\langle {p_{t}} \rangle$ distributions
as described in Sec. III for various collision centralities in Cu+Cu collisions 
at $\sqrt{s_{NN}}$ = 62.4 and 200 GeV.}
\vspace{-0.5cm}
\begin{center}
\begin{tabular}{c|c|c|c|c|c|c|c|c}
\hline
&\multicolumn{4}{l|}{Cu+Cu 62.4 GeV}&\multicolumn{4}{l}{Cu+Cu 200 GeV}\\  [0.1cm]
\cline{2-9} \\ [-0.3cm]
\% cross  & DCA        & $N_{\rm Fit}$     &  DCA        & $N_{\rm Fit}$ & DCA        & $N_{\rm Fit}$     &  DCA        & $N_{\rm Fit}$    \\  
section  & $\mu$ (\%)       & $\mu$ (\%)     & $\sigma$ (\%)       &
$\sigma$ (\%)  & $\mu$ (\%)       & $\mu$ (\%)     & $\sigma$ (\%)       & $\sigma$ (\%)    \\ 
\hline
0-10     & 3.6 &  0.4        & 7.9 & 1.0  & 3.4 &  0.24         &  5.7  &  1.2 \\ [0.1cm]
10-20    & 3.6 &  0.4        & 7.1 & 1.0  & 3.3 &  0.23         &  5.2  &  1.2  \\ [0.1cm]
20-30    & 3.6 &  0.4        & 6.6 & 1.0  & 3.3 &  0.23         & 5.2   &  1.1  \\[0.1cm]
30-40    & 3.6 &  0.4        & 6.2 & 1.0  & 3.3 &  0.20         & 4.7   &  1.2  \\[0.1cm]
40-50    & 3.6 &  0.4        & 6.0 & 1.0  & 3.2 &  0.20         & 4.9   &  1.1  \\[0.1cm]
50-60    & 3.6 &  0.4        & 6.0 & 1.0  & 3.2 &  0.22         & 4.4   &  1.2  \\[0.1cm]
\hline
\end{tabular}
\end{center}
\end{table}

\begin{table}
\caption{\label{tabcorr62_5}
Systematic errors on $p_{t}$ correlations 
as described in Sec. IV for various collision centralities in Cu+Cu collisions at $\sqrt{s_{NN}} =$ 62.4 GeV
and 200 GeV.}
\begin{center}
\begin{tabular}{c|c|c|c|c|c|c|c|c}
\hline
&\multicolumn{4}{l|}{Cu+Cu 62.4 GeV}&\multicolumn{4}{l}{Cu+Cu 200 GeV}\\  [0.1cm]
\cline{2-9} \\ [-0.3cm]
\% cross  & DCA        & $N_{\rm Fit}$     &  Poly.        & Low
$p_t$& DCA    &    $N_{\rm Fit}$     & Poly.        & Low $p_t$    \\  
section  & (\%)       & (\%)     & (\%)       &
(\%)  & (\%)       & (\%)     & (\%)       & (\%)    \\ 
\hline
0-10    &  30 &  1.4      & 1.9      & 7.2     & 16 &  0.05   &1.9  & 22   \\ [0.1cm]
10-20  &  23 &  0.8      & 3.6      & 13.1   & 13 &  0.09   &3.6 & 3.2    \\ [0.1cm]
20-30  &  18 &  0.6      & 1.9      &  3.4    & 13 &  1.1     &1.9 & 12.3 \\ [0.1cm]
30-40  &  17 &  1.0      & 0.004  &  9.0    & 8   &   1.2   &0.004 &  9.7   \\[0.1cm]
40-50  &  19 &  3.0      & 0.009  &  1.0    & 10 &  2.0    &0.009 &8.4   \\ [0.1cm]
50-60  &  20 &  3.6     & 0.009  &  4.0    & 7  &  5.0      &0.009 & 8.3     \\ [0.1cm]
\hline
\end{tabular}
\end{center}
\end{table}

The $p_{t}$ correlations may also include short range correlations such as 
Coulomb interactions and Hanbury Brown-Twiss (HBT) correlations. These correlations
usually dominate among pairs of particles having relative transverse momentum 
less than 100 MeV/$c$. The effect of
these short-range correlations on the measured $p_{t}$ correlations is seen by calculating $p_{t}$ correlations
after removing the pairs of particles with relative momentum ($p_{i} - p_{j}$), less than 
100 MeV/$c$.
The $p_{t}$ correlations are reduced by a maximum of
6\% when short-range correlations are excluded.

The $p_{t}$ correlations also include the contributions from resonance decays and charge ordering.
These correlations are obtained for pairs of 
particles having like (++ or $--$) and unlike ($+-$) charges with respect to 
inclusive charged particles. 
A maximum of 15\% decrease in the correlations is observed for pairs of particles 
with like charges and about 12\% increase is observed for pairs with unlike charges
with respect to the correlations for inclusive charged particle pairs for Cu+Cu 
collisions at $\sqrt{s_{NN}} =$ 62.4 GeV and 200 GeV.

\section{Summary}
We have reported measurements of $p_t$ fluctuations in Cu+Cu
collisions in the STAR detector at RHIC for $\sqrt{s_{NN}}=$ 62.4 and
200 GeV, and compared with Au+Au collisions at the same energies to
investigate the system size dependence. 
The dynamical $p_t$ fluctuations are observed by comparing data to mixed events in both Cu+Cu 
and Au+Au collisions at these two beam energies. Moreover, for similar
mean number of participating nucleons, the $p_t$ fluctuations are observed
to be comparable for the Cu+Cu and Au+Au systems, suggesting that the
system size has little effect. 
In addition, $p_t$ correlation measurements for Cu+Cu collisions at $\sqrt{s_{NN}}=$ 62.4
and 200 GeV are compared with the published Au+Au measurements. 
For both Cu+Cu and Au+Au systems, the $p_t$ correlation decreases with
increasing  $\langle N_{\rm{part}} \rangle$ at both beam energies.
The dilution of the $p_t$ correlations with increasing $\langle
N_{\rm{part}} \rangle$ could be due to the decrease in correlations
that are dominated by pairs of particles that originate from the same
nucleon-nucleon collision. A similar observation is found with other fluctuation and
correlation observables such as $K/\pi$~\cite{k2pi_star} and
net charge~\cite{netch_star}.

The $p_t$ correlations are found to scale with number of participating
nucleon pairs for $\langle N_{\rm{part}} \rangle > 100$ (when the
system size is larger than that of central Cu+Cu collisions). 
This might indicate
the onset of thermalization~\cite{gavin_5},  
the onset of jet quenching~\cite{steph4_5,jetq2_5}, 
or the saturation of transverse flow in central collisions~\cite{tflow_5}.
The square root of the $p_t$ correlations normalized by eventwise
average transverse momentum for Cu+Cu and 
Au+Au collisions is similar for systems with similar $\langle
N_{\rm{part}} \rangle$ and is independent of the beam 
energies studied.

The results described above are compared to predictions from several
relevant model calculations. 
The transport-based URQMD model calculations are found to have a
better quantitative agreement with the measurements compared to models
which incorporate only jet-like correlations as in HIJING. HIJING
gives similar dependence on $\langle N_{\rm{part}} \rangle$, but
under predicts the magnitude. Inclusion of the jet-quenching effect in HIJING leads to a smaller value of the correlations in central collisions. A multiphase transport model calculation incorporating coalescence as a mechanism of particle production also compares well with data for central collisions.

When studying the $p_t$ correlations for different $p_t$ intervals, the correlations
appear to be small and fairly independent
of $p_t$ interval, 
if the lower $p_t$ bound is fixed at 0.15 GeV/$c$ and
the higher $p_t$ bound is progressively
increased up to 0.50 GeV/$c$. This suggests that correlations are weak for low-$p_t$
particles. This low-$p_t$ trend observed in the data is also seen in
URQMD, AMPT, and HIJING models. When the analysis is carried out
keeping the higher $p_t$ bound fixed at 2.0 GeV/$c$ and progressively
decreasing the lower $p_t$ bound to $p_t$ = 0.15 GeV/$c$, the correlation
values in data are found to increase. This suggests that high $p_t$
particles are more correlated with low-$p_t$ particles. The AMPT model shows
a rather similar variation of $p_t$ correlations for different $p_t$ intervals,
as observed in data.
The URQMD model calculations, however, show
no such variations in correlations for the different $p_t$ intervals
with higher $p_t$ bound fixed at 2.0 GeV/$c$.
Finally, it is noted that the HIJING model calculations give $p_t$ correlations
that decrease with a decrease in the lower $p_t$ bound 
for intervals with fixed higher $p_t$ (= 2.0 GeV/$c$) bound,
i.e., 
opposite to what is observed in data. 
Regarding the changes in $p_t$ correlations in different $p_t$
intervals, 
it is found that 
the resultant fluctuations after considering
event-by-event variation in the slope of the $p_t$ spectra for different 
$p_t$ bins are all of similar order. 

The variation of $p_t$ correlation with the change in the accepted range
of pseudorapidity and azimuthal angle of the produced particles, are
also shown.
The correlation values increase when the $\eta$- and
the $\phi$-acceptance are reduced.

\section{Acknowledgements}
We thank the RHIC Operations Group and RCF at BNL, the NERSC Center at
LBNL and the Open Science Grid consortium for providing resources and
support. This work was supported in part by the Offices of NP and HEP
within the U.S. DOE Office of Science, the U.S. NSF, the Sloan
Foundation, CNRS/IN2P3, FAPESP CNPq of Brazil, Ministry of Education
and Science of the Russian Federation, NNSFC, CAS, MoST, and MoE of
China, GA and MSMT of the Czech Republic, FOM and NWO of the
Netherlands, DAE, DST, and CSIR of India, the Polish Ministry of
Science and Higher Education, the National Research Foundation
(NRF-2012004024), the Ministry of Science, Education and Sports of the Republic of Croatia, and RosAtom of Russia.

\end{document}